\documentclass[aps, reprint,noshowpacs,superscriptaddress]{revtex4-2}

\usepackage{graphicx}
\usepackage{gensymb}
\usepackage{amsmath}
\usepackage{braket}
\usepackage{lipsum}
\usepackage[colorlinks,linkcolor=black,citecolor=black,urlcolor=black,filecolor=black]{hyperref}
\usepackage{wrapfig}
\usepackage{CJKutf8}
\usepackage[T1]{fontenc}
\usepackage{titlesec}
\usepackage{color,soul}
\usepackage[normalem]{ulem} 
\usepackage{xcolor}         

\usepackage{lineno}

\begin{document}

\newcommand\numberthis{\addtocounter{equation}{1}\tag{\theequation}}

\newcommand{\titlesize}{\fontsize{13pt}{15pt}\selectfont}


\title{\titlesize A Surface-Scaffolded Molecular Qubit}

\author{Tian-Xing Zheng}
\thanks{These authors contributed equally to this work.}
\affiliation{Pritzker School of Molecular Engineering, University of Chicago, Chicago, IL 60637, USA}
\affiliation{Department of Physics, University of Chicago, Chicago, IL 60637, USA}

\author{M. Iqbal Bakti Utama}
\thanks{These authors contributed equally to this work.}
\affiliation{Department of Materials Science and Engineering, Northwestern University, Evanston, IL 60208, USA}

\author{Xingyu Gao}
\affiliation{Pritzker School of Molecular Engineering, University of Chicago, Chicago, IL 60637, USA}

\author{Moumita Kar}
\affiliation{Department of Chemistry, Northwestern University, Evanston, IL 60208, USA}

\author{Xiaofei Yu}
\affiliation{Pritzker School of Molecular Engineering, University of Chicago, Chicago, IL 60637, USA}
\affiliation{Department of Physics, University of Chicago, Chicago, IL 60637, USA}

\author{Sungsu Kang}
\affiliation{James Franck Institute, University of Chicago, Chicago, IL 60637, USA}
\affiliation{Department of Chemistry, University of Chicago, Chicago, IL 60637, USA}

\author{Hanyan Cai}
\affiliation{Pritzker School of Molecular Engineering, University of Chicago, Chicago, IL 60637, USA}
\affiliation{Department of Physics, University of Chicago, Chicago, IL 60637, USA}

\author{Tengyang Ruan}
\affiliation{Pritzker School of Molecular Engineering, University of Chicago, Chicago, IL 60637, USA}

\author{David Ovetsky}
\affiliation{Pritzker School of Molecular Engineering, University of Chicago, Chicago, IL 60637, USA}

\author{Uri Zvi}
\affiliation{Pritzker School of Molecular Engineering, University of Chicago, Chicago, IL 60637, USA}

\author{Guanming Lao}
\affiliation{Pritzker School of Molecular Engineering, University of Chicago, Chicago, IL 60637, USA}

\author{\begin{CJK}{UTF8}{gbsn}Yu-Xin Wang (王语馨)\end{CJK}}
\affiliation{Joint Center for Quantum Information and Computer Science, University of Maryland, College Park, MD, 20742, USA}

\author{Omri Raz}
\affiliation{Pritzker School of Molecular Engineering, University of Chicago, Chicago, IL 60637, USA}

\author{Sanskriti Chitransh}
\affiliation{Pritzker School of Molecular Engineering, University of Chicago, Chicago, IL 60637, USA}
\affiliation{Department of Physics, University of Chicago, Chicago, IL 60637, USA}

\author{Grant T. Smith}
\affiliation{Pritzker School of Molecular Engineering, University of Chicago, Chicago, IL 60637, USA}

\author{Leah R. Weiss}
\affiliation{Pritzker School of Molecular Engineering, University of Chicago, Chicago, IL 60637, USA}

\author{Magdalena H. Czyz}
\affiliation{Department of Chemistry, Northwestern University, Evanston, IL 60208, USA}

\author{Shengsong Yang}
\affiliation{James Franck Institute, University of Chicago, Chicago, IL 60637, USA}
\affiliation{Department of Chemistry, University of Chicago, Chicago, IL 60637, USA}

\author{Alex J. Fairhall}
\affiliation{Department of Chemistry, University of Wisconsin–Madison, Madison, WI 53706, USA}

\author{Kenji Watanabe}
\affiliation{Research Center for Functional Materials, National Institute for Materials Science, 1-1 Namiki, Tsukuba 305-0044, Japan}

\author{Takashi Taniguchi}
\affiliation{Research Center for Materials Nanoarchitectonics, National Institute for Materials Science, 1-20-1 Namiki, Tsukuba 305-0044, Japan}

\author{David D. Awschalom}
\affiliation{Pritzker School of Molecular Engineering, University of Chicago, Chicago, IL 60637, USA}
\affiliation{Department of Physics, University of Chicago, Chicago, IL 60637, USA}
\affiliation{Center for Molecular Engineering and Materials Science Division, Argonne National Laboratory, Lemont, IL 60439, USA}

\author{A. Paul Alivisatos}
\affiliation{Pritzker School of Molecular Engineering, University of Chicago, Chicago, IL 60637, USA}
\affiliation{James Franck Institute, University of Chicago, Chicago, IL 60637, USA}
\affiliation{Department of Chemistry, University of Chicago, Chicago, IL 60637, USA}

\author{Randall H. Goldsmith}
\affiliation{Department of Chemistry, University of Wisconsin–Madison, Madison, WI 53706, USA}

\author{George C. Schatz}
\affiliation{Department of Chemistry, Northwestern University, Evanston, IL 60208, USA}

\author{Mark C. Hersam}
\email{m-hersam@northwestern.edu}
\affiliation{Department of Materials Science and Engineering, Northwestern University, Evanston, IL 60208, USA}
\affiliation{Department of Chemistry, Northwestern University, Evanston, IL 60208, USA}
\affiliation{Department of Electrical and Computer Engineering, Northwestern University, Evanston, IL 60208, USA}

\author{Peter C. Maurer}
\email{pmaurer@uchicago.edu}
\affiliation{Pritzker School of Molecular Engineering, University of Chicago, Chicago, IL 60637, USA}
\affiliation{Center for Molecular Engineering and Materials Science Division, Argonne National Laboratory, Lemont, IL 60439, USA}
\affiliation{CZ Biohub Chicago, LLC, Chicago, IL 60642, USA}

\begin{abstract}
{\normalsize  
Fluorescent spin qubits are central building blocks of quantum technologies.
Placing these qubits at surfaces maximizes coupling to nearby spins and fields, enabling nanoscale sensing and facilitating integration with photonic and superconducting devices.
However, reducing the dimensions or size of established qubit systems without sacrificing the qubit performance or degrading the coherence lifetime remains challenging.
Here, we introduce a surface molecular qubit formed by pentacene molecules scaffolded on a two-dimensional (2D) material, hexagonal boron nitride (hBN).
The qubit exhibits stable fluorescence and optically detected magnetic resonance (ODMR) from cryogenic to ambient conditions.
With fully deuterated pentacene, the Hahn-echo coherence reaches 22~$\mu$s and further extends to 214~$\mu$s under dynamical decoupling, outperforming state-of-the-art shallow NV centers in diamond, despite being positioned directly on the surface.
We map the local spin environment, resolving couplings to nearby nuclear and electron spins that can serve as auxiliary quantum resources.
This platform combines true surface integration, long qubit coherence, and scalable fabrication, opening routes to quantum sensing, quantum simulation, and hybrid quantum devices. It also paves the way for a broader family of 2D material–supported molecular qubits.
}
\end{abstract}

\maketitle

Exploring physical systems in different dimensionalities often unveils novel opportunities, especially for quantum systems that are inherently nanometer-scale. Among various platforms, optically addressable spin qubits have emerged as a powerful platform for quantum sensing~\cite{degen2017quantum,aslam2023quantum,yu2021molecular}, simulation~\cite{cai2013large,randall2021many,zu2021emergent}, and information processing~\cite{awschalom2018quantum} due to their long coherence times and robust spin–optical readout. A long-sought goal has been to position such qubits at surfaces, where proximity maximizes dipolar coupling to nearby spins and fields, enables nanoscale sensing, and allows direct integration with photonic and superconducting devices.

Yet, placing qubits close to surfaces without degrading their coherence or charge stability has proven exceptionally challenging. Solid-state defect spins in bulk crystals~\cite{wolfowicz2021quantum} can be engineered near surfaces but typically suffer from surface-induced noise~\cite{rosskopf2014investigation}, charge instability~\cite{bluvstein2019identifying}, and reduced coherence, often requiring complex surface treatments~\cite{sangtawesin2019origins}. 
Trapped ions~\cite{leibfried2003quantum} and neutral atoms~\cite{ludlow2015optical} in vacuum exhibit long coherence times, but are poorly suited for chemical and biological sensing applications and require substantial experimental overhead.
\begin{figure*}[t]
\includegraphics[width=0.95\textwidth]{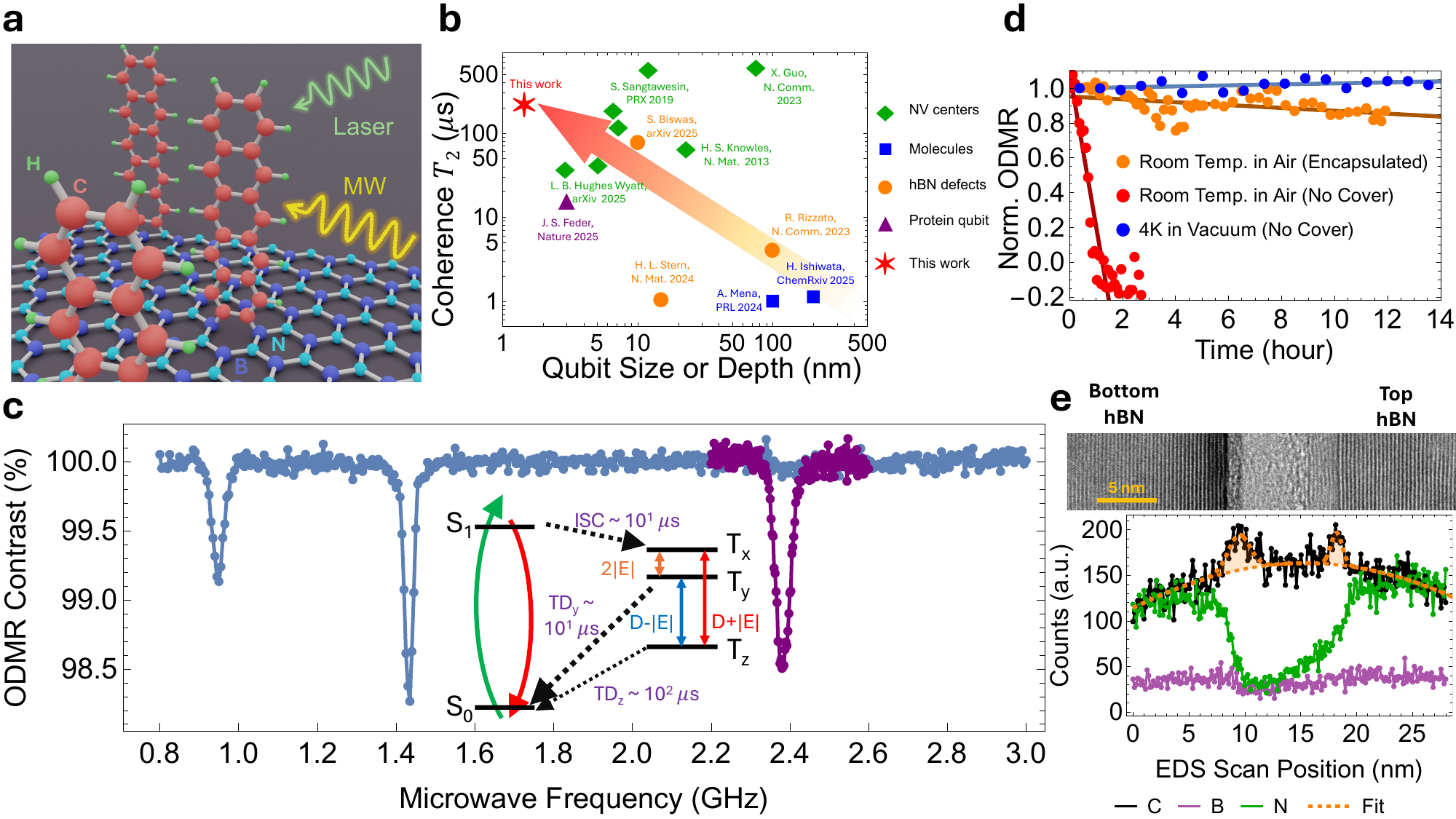}
\caption{\textbf{Photophysics and ODMR of the hBN-scaffolded pentacene.} 
\textbf{a,} Schematic of the pentacene molecular qubit on the surface of hBN. The qubit is excited and read out by an off-resonance green laser ($520$~nm), and its spin levels are controlled by microwaves through our home-built confocal microscopes (see Methods). 
\textbf{b,} The $T_2$ coherence time versus qubit size or depth for optically-addressable spin qubits including: shallow NV centers~\cite{sangtawesin2019origins,wyatt2025creation,knowles2014observing,guo2024direct}, 
molecules~\cite{mena2024room,ishiwata2025molecular}, hBN defects~\cite{rizzato2023extending,stern2024quantum,biswas2025quantum}, protein qubits~\cite{feder2025fluorescent}, and hBN-scaffolded pentacene (this work). The arrow indicates the preferred qubit coherence and size for quantum applications. The longest demonstrated $T_2$ coherence times for different qubit platforms are plotted, regardless of measurement conditions. The $T_2$ reported in this work is obtained using a dynamical decoupling sequence (Fig.~\ref{fig3}f), and the qubit size is defined by the longest molecular dimension of pentacene ($\sim$1.5 nm).
\textbf{c,} ODMR spectrum measured at 4~K in vacuum. Purple: 
The signal at $2.38$~GHz emerges with $\pi$-pulse at the $1.43$~GHz transition during initialization and before the laser readout. The ODMR contrast is defined as $C = I_{\mathrm{MW,on}}/I_{\mathrm{MW,off}}$. Inset: Energy-level diagram of the pentacene molecule, containing a bright singlet manifold $\{S_0, S_1\}$ and a dark triplet manifold $\{T_x, T_y, T_z\}$. 
\textbf{d,} ODMR contrast monitored under continuous-wave laser excitation ($\sim13~$kW/cm$^2$). The fitted half-life of the hBN-scaffolded pentacene qubit at ambient conditions is 37 $\pm$ 3 min, and no bleaching has been observed for qubits at 4 K in vacuum after more than six months. The stability of the pentacene qubit in ambient conditions is improved significantly by hBN encapsulation with half-life of 58 $\pm$ 17 hours.
\textbf{e,} Top: TEM image of the hBN surface obtained from a 100 nm slice cut by focused ion beam (FIB) from a hBN-encapsulated pentacene sample. Bottom: EDS line-scan results showing the distribution of carbon, boron, and nitrogen elements along the vertical direction of the hBN surface. The carbon signal is fitted with two Gaussian components having FWHMs of 2.28 nm and 1.21 nm, respectively.
}
\label{fig1}
\end{figure*}
Molecular spin qubits provide chemical tunability and well-defined electronic structures, but so far have been realized exclusively within crystalline host matrices~\cite{wrachtrup1993optical,kohler1993magnetic,bayliss2020optically,wang2019turning}. 
Fluorescent protein qubits offer diverse opportunities for sensing within biological systems, but their coherence times are limited and they suffer from photobleaching~\cite{feder2025fluorescent}. 
Two-dimensional (2D) materials provide a natural scaffold for surface qubits with controlled orientation, deterministic placement, and seamless compatibility with van der Waals heterostructures. 
However, intrinsic spin defects identified in hexagonal boron nitride (hBN) display limited spin coherence times and faces challenges when it comes to placing these defects at the sufaces~\cite{rizzato2023extending,stern2024quantum,gao2025single,biswas2025quantum}. 
A system that preserves spin coherence while residing directly on a surface would therefore combine the desirable features of surface accessibility and device integration~\cite{kavand2025general}.

Here, we introduce a surface molecular qubit platform based on pentacene molecules scaffolded vertically to hBN (Fig.~\ref{fig1}a). This architecture combines the electronic spin qubit functionality of molecular systems with the integration advantages of 2D materials. The small ensemble of pentacene molecules exhibit stable fluorescence and ODMR from cryogenic to ambient conditions (Extended Data Fig.~4). 
Importantly, as the molecules are positioned on top of, rather than embedded within, the hBN lattice, they remain well separated from the nuclear spin bath and thus retain significantly longer coherence times than intrinsic hBN spin defects. Through isotopic substitution of $^{1}$H with $^{2}$D, the Hahn-echo coherence improves by almost an order of magnitude, from 2.5 $\pm$ 0.1~$\mu$s to 22.4 $\pm$ 0.5~$\mu$s. Incorporating dynamical decoupling and multilevel spin control pushes the coherence time to 214 $\pm$ 19~$\mu$s, outperforming the state-of-the-art coherence of shallow NV centers \cite{sangtawesin2019origins,knowles2014observing,guo2024direct} despite being directly at the surface. As shown in Fig.~\ref{fig1}b, the hBN-scaffolded pentacene achieves the smallest device footprint among diverse optically addressable spin qubits, while still preserving a coherence time competitive with established platforms. We further resolve dipolar interactions with local nuclear and electron spins, highlighting the potential of ancillary quantum resources. Our approach establishes a new pathway for engineering molecular qubits at the surface of 2D hosts, combining long coherence, optical addressability, and facile sample preparation. Beyond hBN, the platform can be generalized to other 2D hosts, paving the way for nanoscale sensing, programmable spin networks in two-dimensional lattices, and hybrid quantum devices that exploit the full versatility of van der Waals heterostructures.

\section*{Molecular qubit hosted by 2D material}
Pentacene is a prototypical molecule featuring an optically addressable spin-triplet state that has been extensively studies in molecular crystals~\cite{wrachtrup1993optical,kohler1993magnetic}. The inset of Fig.~\ref{fig1}c shows the spin energy-level structure, comprising a bright singlet manifold with a ground state ($S_0$), which can be off-resonantly driven to an excited state ($S_1$) and subsequently returns to $S_0$ via fluorescence emission. The excited singlet $S_1$ can also relax to $S_0$ non-radiatively via intersystem crossing (ISC) through the metastable triplet manifold $\{T_x, T_y, T_z\}$, enabling ODMR. This triplet (spin-1) system can be described by the Hamiltonian
\begin{align}
    H = \hbar D(S_z^2-\vec{S}^2/3)+\hbar E (S_x^2 - S_y^2) + \hbar \gamma_e \vec{B}\cdot\vec{S},
    \label{spin-1_hamiltonian}
\end{align}
where $\hbar$ denotes the reduced Planck constant, $D$ and $E$ the zero-field splitting (ZFS) parameters, $\gamma_e=-(2\pi)\times28$~GHz/T the gyromagnetic ratio of an electron spin, $\vec{B}$ the external magnetic field, and $\vec{S}$ the vector of spin-1 matrices. Because the triplet sublevels ${T_x, T_y, T_z}$ exhibit different triplet decay (TD) rates back to the singlet ground state $S_0$, a microwave pulse that swaps the populations of two sublevels causes the molecule’s electronic spins to remain in the triplet manifold for different durations. Consequently, the molecule’s fluorescence rate changes when a microwave field drives transitions between the triplet spin levels. This process underlies the mechanism of ODMR and constitutes a necessary condition for the spin qubit to be optically addressable.

To prepare surface qubits, we use either a solvent-based dip-coating process or a vacuum deposition process of pentacene on hBN flakes that are pre-transferred on a waveguide chip (Methods).   The resulting hBN flakes exhibit bright orange photoluminescence at multiple emission sites appearing across the flakes (Extended Data Fig.~1-3. 
We performed pulsed ODMR at zero magnetic field, resolving all three transitions of the $S=1$ triplet in Fig.~\ref{fig1}c. The protocol begins with a $15~\mu\mathrm{s}$ green laser pulse to initialize the population into metastable triplet states, followed by a frequency-swept microwave drive, and finalized with a short readout pulse after a fixed delay to allow relaxation back to $S_0$ (Extended Data Fig.~6). 
A single microwave pulse reveals the $T_y\!\leftrightarrow\!T_z$ and $T_x\!\leftrightarrow\!T_y$ resonances at $(1.430\pm0.002)$~GHz and $(0.95\pm0.01)$~GHz, respectively. The remaining $T_x\!\leftrightarrow\!T_z$ line at $(2.38\pm0.02)$~GHz appears under a multilevel control sequence: a preparatory $T_y\!\leftrightarrow\!T_z$ $\pi$ pulse to swap populations, a swept probe across $T_x\!\leftrightarrow\!T_z$, and a final $T_y\!\leftrightarrow\!T_z$ pulse for optimal readout. Solving the $S=1$ spin Hamiltonian (Eq.~\ref{spin-1_hamiltonian}) at zero field yields three transition frequencies at $|D+E|$, $|D-E|$, and $|2E|$, which match with the three signals observed in the ODMR spectrum with ZFS parameters of $D = 2\pi \times (1.905 \pm 0.005)$~GHz and $E = -2\pi \times (0.475 \pm 0.005)$~GHz. 
The observed contrast of less than unity reflects optical pumping into a short-lived state with a high ISC rate that results in brighter readout. Resonant microwaves, on the other hand, transfers population into longer-lived, darker states. 
The need to employ a multi-level control sequence to efficiently read out the $T_x\!\leftrightarrow\!T_z$ transition indicates nearly equal optical steady state population of $T_x$ and $T_z$ state. Notably, in negative control experiments, no ODMR signal is observed when pentacene is absent, when oxidized pentacene is used, or when other 2D substrates are employed. These negative control results further strengthen the hypothesis that the observed ODMR signal originates from an interaction between pentacene molecules and the hBN substrate (Supplemental Table~S2).

The photostability of optically addressable molecular qubits is a key requirement for scalable quantum technologies and long-term sensing applications. Although, at cryogenic conditions pentacene in molecular crystals~\cite{wrachtrup1993optical,kohler1993magnetic} and other organic fluorophores~\cite{feder2025fluorescent,navarro2014stable,sartor2023characterization} can possess exceedingly long coherence times, their performance degrades substantially at higher temperatures. At ambient conditions, organic triplet emitters are intrinsically susceptible to photo-induced bleaching via singlet oxygen generation and excited-state reactions~\cite{maliakal2004photochemical, ha2012photophysics}. In contrast to these known limitations, we find that pentacene molecules scaffolded on the surface of hBN exhibit markedly enhanced photostability under ambient conditions (Fig. \ref{fig1}d and Supplemental Fig.~S3). 
Under continuous 520-nm illumination at an intensity of $\sim13~$kW/cm$^2$, our pentacene emitters exhibited a photobleaching-induced ODMR contrast lifetime of 37 $\pm$ 3 min in air. Encapsulation with an additional top hBN layer further suppresses photo-oxidation, resulting in less than 15 $\pm$ 1 \% signal loss after 12 h of continuous excitation. At cryogenic temperatures, no measurable bleaching was detected over several months, highlighting the exceptional chemical and structural protection conferred by the hBN scaffold. 

Fig.~\ref{fig1}e shows a cross-sectional transmission electron microscopy (TEM) image revealing the layered structure of the hBN-encapsulated pentacene molecules. The accompanying energy-dispersive X-ray spectroscopy (EDS) map reveals a maximum pentacene layer thickness of approximately 2.3~nm (see details in the Supplemental Fig.~S1-S2).

\begin{figure}[t]
\includegraphics[width=0.5\textwidth]{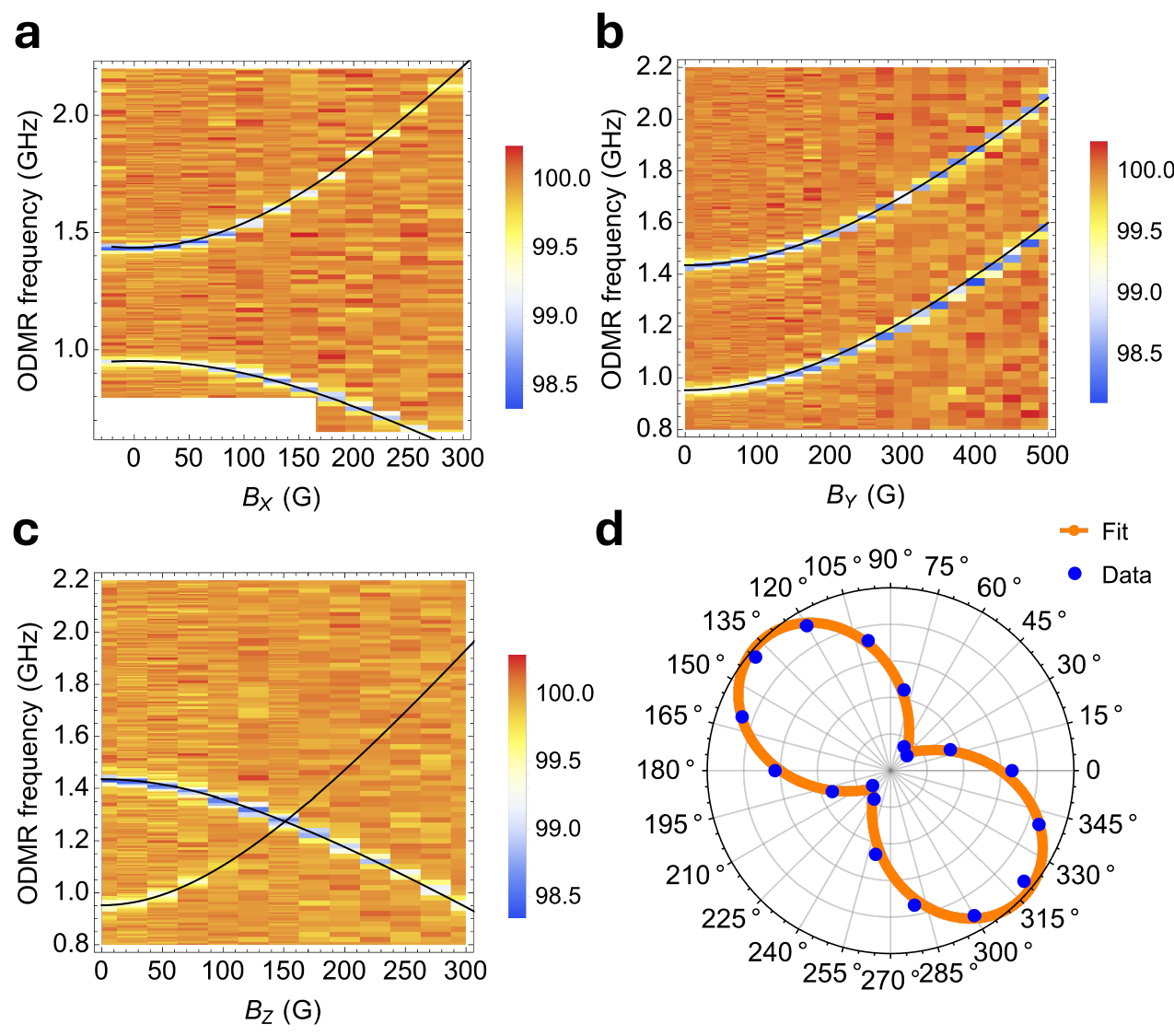}
\caption{\textbf{Orientation of the qubits on the hBN scaffold.} Dependence of ODMR signals on the external magnetic field applied perpendicular to (\textbf{a}) and within (\textbf{b} \& \textbf{c}) the hBN plane. Solid black curves represent the simulated pentacene molecules' ODMR spectrum based on the $D$ and $E$ values measured in Fig.~\ref{fig1}c and applying the magnetic field along the $X$ (\textbf{a}), $Y$ (\textbf{b}) and $Z$ (\textbf{c}) directions of the spin triplet (Eq.~\ref{spin-1_hamiltonian}). 
%
%
\textbf{d,} Fluorescence intensity (arbitrary unit) of the hBN-scaffolded pentacene versus the excitation laser's polarization. 
%
%
}
\label{fig2}
\end{figure}

\section*{
Orientation of hBN-scaffolded pentacene qubit} 
To investigate the nature of the pentacene qubit on hBN, we performed ODMR under magnetic fields applied along three orthogonal axes. Because of the large zero-field splitting parameters $D$ and $E$ arising from the low molecular symmetry of pentacene, the field dependence along these axes differs strongly (Fig.~\ref{fig2}, a to c), allowing for an unambiguous assignment of the intrinsic spin orientations. The experimental data are well reproduced by numerical simulations of an $S=1$ Hamiltonian that has $S_x$ perpendicular, and $S_y$ and $S_z$ parallel to the hBN surface. Interestingly, across all ODMR-active sites, we find that this spin orientation remains conserved (an example of such a configuration is illustrated in Fig.~\ref{fig1}a). For a full simulation of the B-field dependent ODMR spectrum, including ODMR contrast, we refer to Extended Data Fig.~5. 

The orientation of optical transition dipole ($\vec{d}$ between $S_0 \leftrightarrow S_1$) can be characterized by the fluorescence intensity ($I$) of the qubit when the the E-field's direction of the excitation laser is changing, since $I\propto\vec{d}\cdot\vec{E}$. As we rotate the linear polarization of the excitation laser, the fluorescence exhibits a clear anisotropy $I(\theta)\!\propto\!\cos^2(\theta-\theta_0)$ (Fig.~\ref{fig2}d), where $\theta_0$ denotes the orientation of the dipole's in-plane projection. 

It is worth pointing out that the aforementioned fluorescence and ODMR signals do not arise from a single pentacene molecule. Based on the clean ODMR spectrum (Fig.~\ref{fig2}a–c) and the single optical dipole (Fig.~\ref{fig2}d), we infer that the local structure of the qubits is an ordered array of pentacene molecules on hBN with a lateral size of $\sim$50~nm (Supplemental Fig.~S2)
, with all spin active pentacene molecules facing to the same direction, as illustrated in Fig.~\ref{fig1}a.


To understand the molecular geometry responsible for the observed ODMR signal, we performed density functional theory (DFT) calculations. We found that when boron and/or nitrogen vacancies are present, DFT calculations predict that pentacene can form covalent bonds with the defect sites, stabilizing a near-upright orientation on the hBN surface (Supplemental Fig.~S12-S18). 
Consistent with our ODMR measurements, the DFT calculations predict that these ``standing'' pentacene molecules have a $S_z$ ($D$ axis) parallel to the hBN plane. This defect-mediated configuration provides both the mechanical isolation and local distortion required for ODMR detection. Ab initio calculations further predict that covalent attachment to B or N vacancies enhances the zero-field splitting parameters, resulting in a significantly increased transverse term $E$ and a modified axial term $D$. These trends align closely with our experimental observations (Supplemental Fig.~S19), 
where hBN-anchored pentacene exhibits substantially increased $D$ and $E$ values compared with NaCl~\cite{sellies2023single} or p-terphenyl~\cite{wrachtrup1993optical,kohler1993magnetic} crystal host. 

This defects-scaffolded ``standing'' pentacene geometry is consistent with the previous studies showing that pentacene molecules adopt a flat-lying geometry on pristine hBN~\cite{amsterdam2020tailoring} and a upright orientation on defective hBN~\cite{gunder2020van}. Additionally, free neighboring molecules tend to aggregate through intermolecular $\pi$-$\pi$ stacking, leading to electronic delocalization and complex spin interactions such as singlet fission~\cite{zimmerman2010singlet,zirzlmeier2015singlet}. The absence of any singlet-fission signatures in our ODMR spectra therefore indicates that the observed signal does not originate from such $\pi$-$\pi$ stacked aggregated ensembles (Supplemental Fig.~S4).

On the other hand, in an isolated planar aromatic molecule (including pentacene), ISC is spin-forbidden due to the high symmetry~\cite{patterson1984intersystem,yu2012spin}. Efficient population transfer through ISC requires local symmetry breaking and vibronic spin–orbit mixing. Consequently, the host environment is essential not only for immobilizing and orienting the molecules but also for providing the electronic asymmetry and state mixing required for rapid ISC and optical spin initialization—an approach exploited in bulk crystalline hosts~\cite{kryschi1992vibronically}. The presence of defect sites on the hBN surface may enable pentacene molecules to chemically bond to the lattice as the most energetically stable configuration. This scaffolding anchors the molecules in a near-upright orientation and supplies the local symmetry breaking required for efficient ISC.

\begin{figure*}[t]
\includegraphics[width=0.99\textwidth]{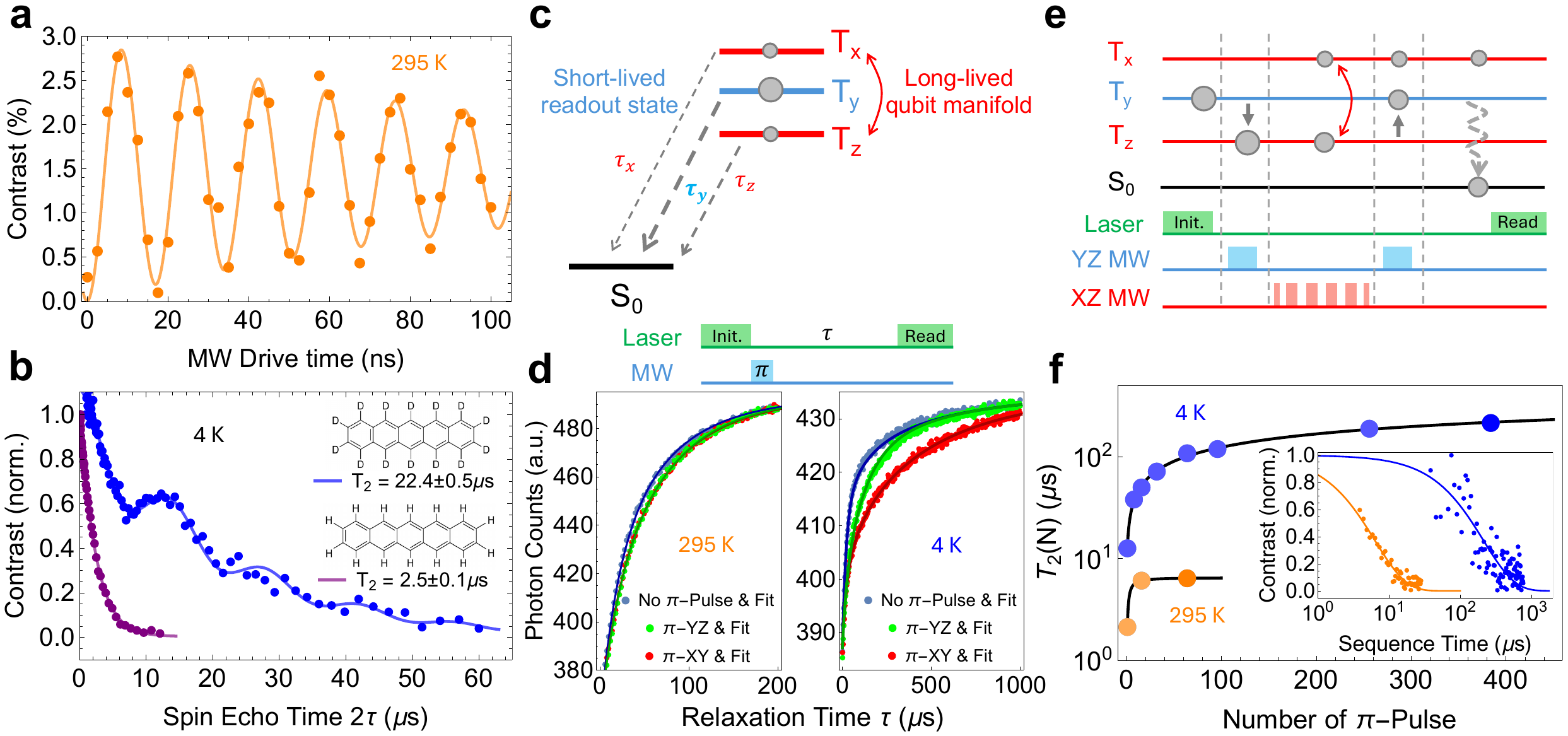}
\caption{\textbf{Coherent control of the qubits.} \textbf{a,} Room temperature Rabi oscillation, with $\Omega_R = 2\pi\times(58.9\pm0.1)$~MHz, of the hBN-scaffolded pentacene qubit driven by microwave at the $T_y\leftrightarrow T_z$ transition. 
\textbf{b,} 
Spin echo $T_2$ of the pentacene-h$_{14}$ (purple) and d$_{14}$ (blue) qubits scaffolded by hBN at 4~K. The h$_{14}$'s data is fitted to $\exp[-(t/T_2)^\nu]$ with $\nu=1.05\pm0.05$, and the d$_{14}$ is fitted to a phenomenological model $\exp[-(t/T_2)^\nu][a-b\sin^2(\omega t/4)]$~\cite{childress2006coherent} with $\nu=1.10\pm0.05$ and $\omega=2\pi\times140.2$ kHz. Inset: structure of pentacene-h$_{14}$ and pentacene-d$_{14}$ molecules. 
%
\textbf{c,} Schematic of multi-level control. The most populated, short-lived spin level $T_y$ is used for readout, while two long-lived spin levels serve as the qubit states.
\textbf{d,} The populations and lifetimes of the triplet states are measured by initializing the molecule into its steady state through laser excitation and subsequently monitoring its fluorescence ($S_0$ population) after a relaxation time $\tau$. The data is fitted to triple exponential decay with the $T_{x},T_{y},T_{z}$ states' population and lifetimes as parameters (see results in Tab.~\ref{tab:pentacene-kinetics}).
\textbf{e,} Multi-level coherent control sequence of the hBN-scaffolded pentacene qubit.
\textbf{f,} $T_2$ coherence time of the long-lived $T_x$–$T_z$ transition measured using XY8 (4~K) and CPMG (295~K) dynamical decoupling sequences. The longest $T_2$ reaches $214 \pm 19~\mu$s at 4~K and $6.4 \pm 0.4~\mu$s at ambient conditions. Data fitted with $T_2(N)^{-1} = (T_2(1)\cdot N^\nu)^{-1}+(2T_{1\rho})^{-1}$, where $N$ denotes the number of $\pi$-pulses, $\nu = 0.53 \pm 0.02, 1.23 \pm 0.1$, $T_{1\rho} = 405 \pm 156, 3.2 \pm 0.1~\mu$s for 4~K and ambient conditions respectively.  Inset: Coherence time data of the longest measured $T_2$ see Supplemental Fig.~S7 for full data). 
}
\label{fig3}
\end{figure*}

\section*{Coherence of hBN-scaffolded pentacene qubit}
Coherent control of the hBN-scaffolded pentacene is achieved by applying microwave pulses on resonance with a selected spin transitions. Sweeping the microwave duration yields distinct Rabi oscillations on the $T_y\!\leftrightarrow\!T_z$ transition (Fig.~\ref{fig3}a). The decay of the Rabi contrast is consistent with its $T^*_2 = 195\pm18$~ns which reflects the local disorder of the pentacene molecules (Extended Data Fig.~7). 
Spin coherence $T_2$ is then characterized by a Hahn-echo sequence. At $T=4$~K and $B=0$, our pentacene samples (pentacene-h$_{14}$) exhibit $T_2 = 2.5 \pm 0.1~\mu\mathrm{s}$, whereas isotopic substitution of the hydrogen with deuterium (pentacene-d$_{14}$) extends $T_2$ to $22.4 \pm 0.5~\mu\mathrm{s}$ (Fig.~\ref{fig3}b). Additionally, the $140.2$ kHz echo-signal modulation of the pentacene-d$_{14}$ qubit agrees with the nitrogen-14 nuclear spin quadrupole moment in hBN~\cite{lovchinsky2017magnetic}. Under more advanced dynamical decoupling, both pentacene-h${14}$ and pentacene-d${14}$ exhibit modest increases in coherence times, and saturate at similar values of $33.9 \pm 2.7~\mu$s and $35.9 \pm 5.9~\mu$s, respectively (see Supplemental Fig.~S5,~S6). 
The fact that, under dynamical decoupling, the terminal qubit coherence (encoded in the $T_y$--$T_z$ manifold) does not depend on the nuclear-bath composition indicates that a mechanism distinct from nuclear spin noise limits coherence.

\begin{table}[h]
\centering
\caption{Kinetic parameters of pentacene molecular qubit on hBN.}
\label{tab:pentacene-kinetics}
\begin{tabular}{l|c|c}
\hline\hline
Triplet
  & \shortstack{Lifetime ($\mu\mathrm{s}$)}
  & \shortstack{Steady state population (\%)} \\
\hline
$T_x$ (Room T.) & $73 \pm 3$   & $30.5 \pm 0.6$ \\
$T_y$ (Room T.) & $18.9 \pm 0.4$   & $41.6 \pm 0.3$ \\
$T_z$ (Room T.) & $61 \pm 2$ & $27.9 \pm 0.5$ \\
\hline
$T_x$ (4~K) & $514 \pm 7$   & $26.3 \pm 0.5$ \\
$T_y$ (4~K) & $21.2 \pm 0.6$   & $53.8 \pm 0.4$ \\
$T_z$ (4~K) & $111 \pm 2$ & $19.9 \pm 0.5$ \\
\hline
\multicolumn{3}{c}{$S_1 \to$ Triplet ISC time scale $\sim 10^1 ~\mu s$} \\
\hline\hline
\end{tabular}
\end{table}

\begin{figure*}
\includegraphics[width=0.99\textwidth]{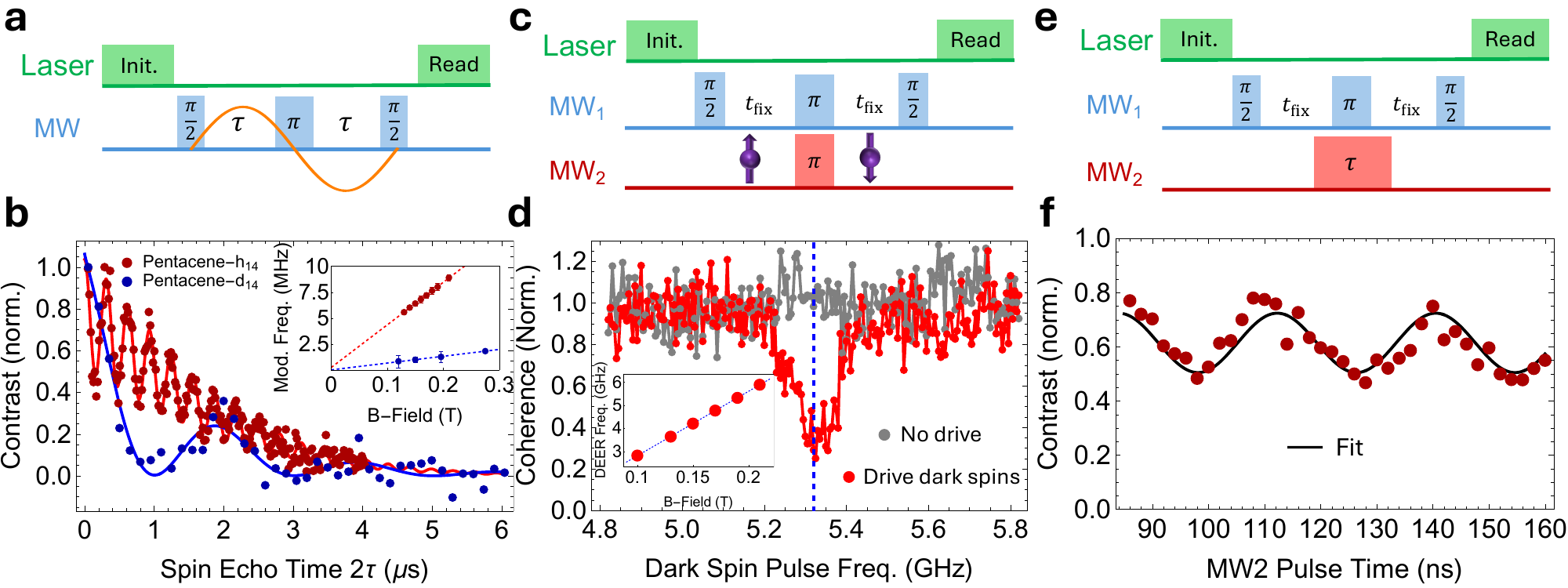}
\caption{\textbf{Detecting and controlling the spin environment}
\textbf{a,} An AC magnetic field signal (orange curve) can be detected when its period matches the duration ($2\tau$) of the spin-echo sequence. In this case, the qubit coherence is suppressed due to the random phase of the AC signal.
\textbf{b,} Spin echo measured at 1500 G aligned to the molecule's Z-axis for hBN-scaffolded pentacene-h$_{14}$ and pentacene-d$_{14}$. The modulations appear at $(2k+1)/f_{\text{AC}} = 2\tau, k=0,1,2,3...$ Inset: modulation frequencies of the spin echo signal under different B-fields (see Supplemental Fig.~S10 for full data). 
\textbf{c,} Pulse sequence of double electron-electron resonance (DEER). The $\pi$-pulse (red) flips the electron spins simultaneously with the refocus $\pi$-pulse (blue) in spin echo, and induces extra decoherence on the fluorescent qubit. 
\textbf{d,} DEER spectrum measured at 1900~G with $t_{\text{fix}}=500$~ns. Blue dashed line: Zeeman splitting of a $g=2$ electron. Inset: frequencies of the DEER signal versus the magnetic field (see Supplemental Fig.~S11 for full data). 
\textbf{e,} Pulse sequence for coherently driving the dark electron spin that couples to the hBN-scaffolded pentacene qubit. 
\textbf{f,} Rabi oscillation, with $\Omega_R = 2\pi\times(35.4\pm0.8)$~MHz, of the dark electron spins measured by the fluorescent pentacene qubit with. The reduced contrast is induced by the local disorder of the dark spins.
}
\label{fig4}
\end{figure*}

To investigate the underlying mechanisms we measured time-domain population dynamics in the triplet state. Using all-optical and microwave-assisted $T_1$ protocols (Fig.~\ref{fig3}d), we observe a short triplet lifetime of approximately 20~$\mu$s for the $T_y$ and a significantly longer triplet lifetime for the $T_x$ and $T_z$ state. For a complete list of the measured triplet lifetimes and triplet populations, see Tab.~\ref{tab:pentacene-kinetics}. Thus, any superposition that includes $T_y$ is ultimately bounded by its lifetime, consistent with the observed saturation of $T_2$ under dynamical decoupling (Supplemental~S6).

We therefore adopt a ``shelf-and-probe'' control scheme that encodes the qubit in the long-lived $\{T_x,T_z\}$ manifold and reserves $T_y$ as a bright, fast initialization/readout ancilla (Fig.~\ref{fig3}c)~\cite{mena2024room}. The modified pulse sequence is shown in Fig.~\ref{fig3}e. After laser initialization, a $T_y\!\leftrightarrow\!T_z$ $\pi$-pulse transfers population from $T_y$ to $T_z$, after which coherent evolution is carried out within the long-lived $\{T_x,T_z\}$ manifold. A selective $\pi$ pulse ($T_x\!\leftrightarrow\!T_y$ or $T_z\!\leftrightarrow\!T_y$) then maps the addressed eigenstate to $T_y$, which rapidly returns to $S_0$ for optical readout. This architecture allows us to maintain coherence well beyond the limits set by the $T_y$'s decay rate. In the deuterated sample, the $T_x\!-\!T_z$ coherence surpasses the $T_y$ bound under XY8-$n$ and approaches the $T_x, T_z$ lifetime limit, reaching $214 \pm 19~\mu\mathrm{s}$ (Fig.~\ref{fig3}f).

We emphasize that the hBN-scaffolded pentacene qubits also exhibit spin polarization (Tab.~\ref{tab:pentacene-kinetics}) and coherence (Fig.~\ref{fig3}f) under ambient conditions. In this case, the $T_2$ coherence time reaches $6.4 \pm 0.4~\mu$s using a CPMG dynamical decoupling sequence at room temperature in air. Taken together, these findings demonstrate that surface molecular qubits can simultaneously achieve long coherence and efficient optical initialization and readout by exploiting the intrinsic lifetime hierarchy of the triplet manifold.

\section*{Accessing environmental quantum resources} 

Having established coherent control of the pentacene qubit, we next use its coherence to probe the surrounding spin environment. These nearby electron or nuclear spins are not directly optically addressable, but can serve as ancillary quantum resources. We first access nearby nuclear ancillae by operating the qubit as a nanoscale nuclear magnetic resonance (NMR) sensor under magnetic field of 0.1 to 0.3~T (Supplemental Fig.~S8,~S9). 
With a static field applied along the molecular spin's $Z$ axis, the precessing proton magnetization generates an AC magnetic field that modulates the probe’s spin-echo signal (Fig.~\ref{fig4}a and b).  To overcome the spectral linewidth imposed by the finite $T_2$, we employ correlation spectroscopy, which extends the resolution to the spin lifetime limit (Extended Data Fig.~8). 
The extracted NMR frequency increases linearly with field, yielding a slope of $41.0 \pm 0.3$~MHz/T and $6.6 \pm 0.2$~MHz/T (Fig.~\ref{fig4}b inset), in good agreement with the proton and deuterium gyromagnetic ratios, $42.58$~MHz/T and $6.54$~MHz/T respectively.  These nuclear spins could potentially be selectively addressed and utilized as auxiliary quantum registers~\cite{van2012decoherence,taminiau2014universal}.

We also use double electron–electron resonance (DEER) to probe the proximal, optically dark electron spins~\cite{grotz2011sensing,mamin2012detecting}. When a secondary microwave tone is swept during the probe’s Hahn-echo sequence, the echo contrast collapses nearly to zero at a distinct resonance (Fig.~\ref{fig4}c and d), evidencing dipolar coupling to an ensemble of dark electron spins. Mapping the DEER resonance frequency versus applied field yields a gyromagnetic ratio $\gamma_{\mathrm{dark}}=(2\pi)\times(27.9\pm0.5)$~GHz/T, consistent with an electron spin with $g=2.00\pm0.04$ (Fig.~\ref{fig4}d inset). Neither isolated pentacene nor pristine hBN hosts unpaired electrons, so the robust DEER signal supports an assignment to defect-related spins at hBN vacancies, nanopores~\cite{dai2023evolution}, or organic radical spins external to the hBN.

Beyond spectral detection, we coherently drive the dark electron spins and monitor their dynamics via the fluorescent qubit through DEER Rabi measurements. Inserting a resonant, variable-duration microwave drive on the dark-spin transition inside the probe’s echo sequence produces oscillations whose frequency increases proportionally in the drive amplitude and vanishes when off-resonant,
demonstrating coherent Rabi rotations of the dark-spin ensemble as read out by the surface molecular qubit. Together, these measurements establish proximal electron spins as controllable ancillary resources that can be addressed and manipulated through the hBN-scaffolded pentacene sensor.

\section*{Discussion}
Our results position hBN-scaffolded pentacene as a new type of optically addressable surface spin qubit operating from cryogenic temperature to ambient conditions. The deuterated sample exhibits a $214 \pm 19~\mu$s spin-coherence time at 4 K under dynamical decoupling, substantially exceeding coherence times reported for native defects in hBN. In fact, our measured coherence times are comparable to the longest values reported for near-surface NV centers in diamond~\cite{sangtawesin2019origins,knowles2014observing,guo2024direct} --- a platform that has benefited from decades of intensive materials optimization --- yet our system achieves this while residing an order of magnitude closer to the surface (Fig.~\ref{fig1}b).
%

It is worth emphasizing that the standing pentacene geometry, where a pentacene molecule forms a covalent bond to an hBN defect site (Fig.~\ref{fig1}a), emerges as the most consistent structural model based on converging experimental and theoretical evidence.
First, ODMR signals are absent in control samples prepared from oxidized pentacene or deposited on substrates other than hBN (Supplemental Table~S2), 
indicating that both intact pentacene and hBN defects are required for qubit formation. The clear observation of proton and deuteron NMR signals further confirms that these ODMR centers originate from pentacene molecules rather than intrinsic hBN defects.
Second, field-dependent ODMR spectroscopy 
show that the axial part D of the zero-field splitting is oriented parallel to the hBN surface. Such a robust preferred orientation suggests a stabilizing interaction, plausibly explained by a covalent bonding configuration that locks the molecule into an upright geometry.
Third, first-principles simulations support this picture, predicting that pentacene forms covalent bonds with hBN defects and relaxes into a ``standing-up’’ configuration. The experimentally measured D and E parameters agree far better with DFT predictions for pentacene conjugated to hBN defects than with reported values for pentacene in conventional host lattices (Supplemental Fig.~S19). 
Fourth, ODMR-active sites appear preferentially near edges, wrinkles, and cracks, regions known to contain elevated defect densities, reinforcing the link between defect availability and qubit formation. 
In addition, DEER measurements reveal strong coupling to nearby dark spins, consistent with DFT predictions that the pentacene–defect bonding configuration produces delocalized unpaired electrons in the surrounding lattice.
Finally, TEM and EDS measurements reveal a pentacene layer with thicknesses down to $\sim$2.3~nm, fully compatible with a model of upright pentacene molecules anchored by hBN surface defects.

\section*{Outlook}
Since the molecule spin resides directly on the surface of a 2D host, this platform is naturally suited to probe nanoscale phenomena that are difficult to access with bulk-embedded defects~\cite{maletinsky2012robust}, including physics in 2D moiré superlattices~\cite{maletinsky2012robust,cao2018unconventional,klein2024imaging} and biological systems~\cite{mateti2018biocompatibility,merlo2018boron}. 
Methodologically, our sample preparation provides a simple and reproducible route, via solution dip-coating or vacuum sublimation, to realize surface qubits without complicated treatments. This approach should be generalizable to other molecule/2D material combinations~\cite{han2021photostable, smit2023sharp}, offering chemical control of ZFS parameters, ISC rates, and optical selection rules. The measurements of the triplet state dynamics with optical readout furnish valuable benchmarks for first-principles theory and are of fundamental interest for molecular spin design.

Looking forward, integrating surface molecular qubits with lithographic defect generation and surface chemistry could enable deterministic positioning, programmable spin networks, and coherent spin–spin coupling across designed lattices. The same scaffolding principle may extend to heterogeneous interfaces, allowing optically addressable molecular qubits to be attached where proximity is crucial. Together, these directions position surface-bound molecular spins as a versatile platform for quantum sensing, simulation, and engineering of hybrid quantum devices.


\section*{\large Acknowledgments}
\noindent We thank Ke Bian, Liang Jiang, Peter Littlewood, Noah Mendelson, Elizabeth Park, Shreya Verma, Xirui Wang and Ruolan Xue for helpful discussions. 

\noindent{\bf Funding:}  The pentacene scaffolding on hBN, sample characterization, coherent control and computation were supported by the Center for Molecular Quantum Transduction, an Energy Frontier Research Center funded by the U.S. Department of Energy, Office of Science, Office of Basic Energy Sciences, under Award No. DE-SC0021314. The cryogenic instrumentation development was supported by the US Department of Energy Office of Science National Quantum Information Science Research Centers as part of the Q-NEXT center. The investigation into the metastable triplet state and related dynamics was supported by the National Science Foundation (NSF) Quantum Leap Challenge Institute QuBBE OMA-2121044. The identification of applications for optically addressable qubits scaffolded on hBN was supported through NSF OSI-2435363. This work was performed, in part, at the National Science Foundation Materials Research Science and Engineering Center at Northwestern University under Award No. DMR- 2308691. This work made use of the GIANTFab core facility at Northwestern University. GIANTFab is supported by the Paula M. Trienens Institute for Sustainability and Energy and the Office of the Vice President for Research at Northwestern. 
K.W. and T.T. acknowledge support from the JSPS KAKENHI (Grant Numbers 21H05233 and 23H02052) , the CREST (JPMJCR24A5), JST and World Premier International Research Center Initiative (WPI), MEXT, Japan. A.J.F and R.H.G~acknowledge support from the fellowship award under contract FA9550-21-F-0003 through the National Defense Science and Engineering Graduate (NDSEG) Fellowship Program, sponsored by the Air Force Research Laboratory (AFRL), the Office of Naval Research (ONR) and the Army Research Office (ARO).
Y.-X.W.~acknowledges support from a QuICS Hartree Postdoctoral Fellowship. 

\noindent{\bf  Author contributions:}
T.-X.Z., X.G., X.Y., H.C., T.R., D.O., U.Z., G.L., O.R. and S.C. performed the ODMR and related measurements. M.I.B.U. and M.H.C. prepared and characterized the sample. M.K. performed the 1st principle simulation. S.K. and S.Y. performed the TEM measurements. T.-X.Z., S.C., G.T.S. and L.R.W. designed and built the low temperature setup. T.-X.Z, M.I.B.U., X.G., M.K., T.R., Y.-X.W., L.R.W. and A.J.F. analyzed the data. K.W. and T.T. synthesized the hBN. D.D.A., A.P.A., R.H.G., G.C.S., M.C.H. and P.C.M. supervised the project. All authors contributed to the
discussions and production of the manuscript.

\noindent{\bf Competing interests:}
T.-X.Z., M.I.B.U., M.C.H. and P.C.M. are inventors on a pending patent application with the USPTO submitted by the University of Chicago and Northwestern University that covers molecular qubits hosted on the surfaces of two-dimensional materials and associated methods.

\noindent{\bf  Data and materials availability:}
The data that support the findings of this study are available from the corresponding author on request.

\noindent {\bf Correspondence and requests for materials} should be addressed to pmaurer@uchicago.edu and m-hersam@northwestern.edu\\




\let\oldaddcontentsline\addcontentsline
\renewcommand{\addcontentsline}[3]{}
\bibliography{library.bib}
\let\addcontentsline\oldaddcontentsline


\end{document}